\documentclass[floatfix,twocolumn,showpacs,preprintnumbers,amsmath,amssymb]{revtex4}
\usepackage{graphicx}
\usepackage{dcolumn}
\usepackage{bm}
\usepackage{subfigure}

\begin{document}

\preprint{}

\title{Is a rubidium cell with long decay time always
useful for generating a non-classical photon pair?}

\author{Qun-Feng Chen}
\email[Electronic address: ]{qfchen@mail.ustc.edu.cn}
\author{Xiao-Song Lu}
\author{Bao-Sen Shi}
\email[Electronic address: ]{drshi@ustc.edu.cn}
\author{Guang-Can Guo}
\affiliation{Key Laboratory of Quantum Information, University of Science and Technology of China, Hefei, 230026, People's Republic of China}


\date{\today}

\begin{abstract}
We experimentally find an interesting and unexpected thing: a
rubidium cell with long decay time can not be used to generate a
non-classical correlated photon pair via the D2 transition of $^{87}$Rb using
four-wave mixing configuration [Opt. Express {\bf 16}, 21708 (2008)]. In this work, we give a detail
theoretical analysis on the EIT of hot $^{87}$Rb with different ground decay time, which shows a probable reason why a rubidium cell with long decay time is not a useful candidate for preparation of a non-classical photon pair via the D2 transition. The simulations  agree well with the
experimental results. We believe our find is very instructive to
such kind of research.
\end{abstract}


\maketitle

It is well known that a rubidium cell with cell filled with buffer
gas or wall paraffin coated can greatly decrease the decay between
the ground states, and in most cases, the decrement of such decay
can greatly improve the performance of the system. The cell with
paraffin coated or buffer gas filled is extensively used in
experiments in atomic field, for example, recently, in the
experiments of the generation of non-classical correlated photon
pairs\cite{Wal:2003:196,Eisaman:PRL:2004:233602,Eisaman:N:2005:837,
manz:040101}. In these works, a non-classical correlated photon pair
is successfully generated using Raman scattering
\cite{Duan:N:2001:413} via the D1 transition of Rb in a cell filled
with buffer gas. Very recently, we prepare non-classical correlated
photon pairs using non-degenerate four-wave mixing in a rubidium
cell\cite{chen:oe:2008,chen:053810}. During the experiments, we find
an interesting and unexpected thing: a normal rubidium cell is a
good candidate for the generation of non-classical non-degenerate
photon pairs using both the D1 and D2 transitions of $^{87}$Rb; On
the contrary, we could not obtained the photon pairs via the D2
transition if a cell coated with paraffin or filled with buffer gas
is used. We try the cells coated with paraffin, and filled with 30
Torr and 8 Torr's neon respectively in experiments, we could observe
the stimulated four-wave mixing in these cells, but can not obtain
the correlated photon pairs. We think this counter-intuitive result
is very probably caused by the small split of the D2 transition of
rubidium combined with the large Doppler broadening.
Electromagnetically induced transparency (EIT) is the key part of
this kind of experiments\cite{balic:183601,Kolchin:PRA:2007:033814}.
However, this combination will make the EIT disappeared, if the
decay between the ground states is ignorable. The disappearance of
EIT makes it impossible to generate a coherence photon. Therefore no
correlated photons can be obtained when a cell with ignorable decay
between the ground states is used. Our theoretical analysis shows
that the decay between the ground states can make the EIT reappear,
which makes the generation of a correlated photon pair available.
The experiment on the EIT effect of the D2 line of $^{87}$Rb with
different kinds of cells supports our calculation. We believe our
find is very instructive to such kind of research.

\begin{figure}[h!]
  \subfigure{\label{fig:level}
  \centering
  \includegraphics[width=4cm]{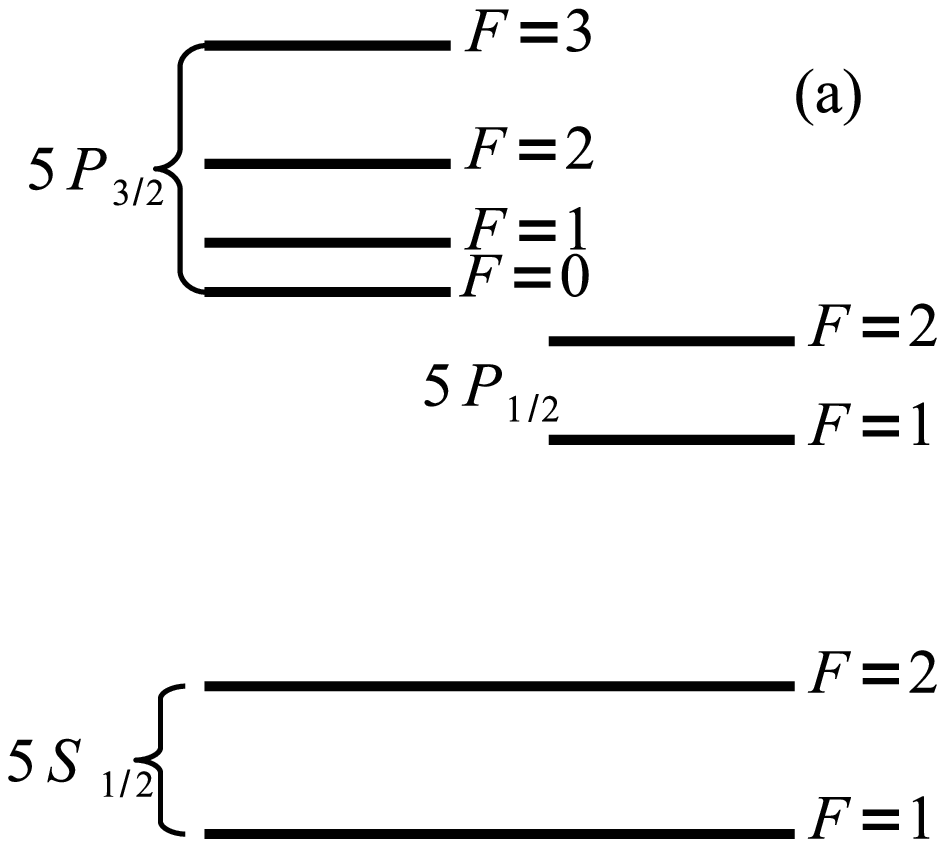}}
  \subfigure{\label{fig:setup}\centering
  \includegraphics[width=4cm]{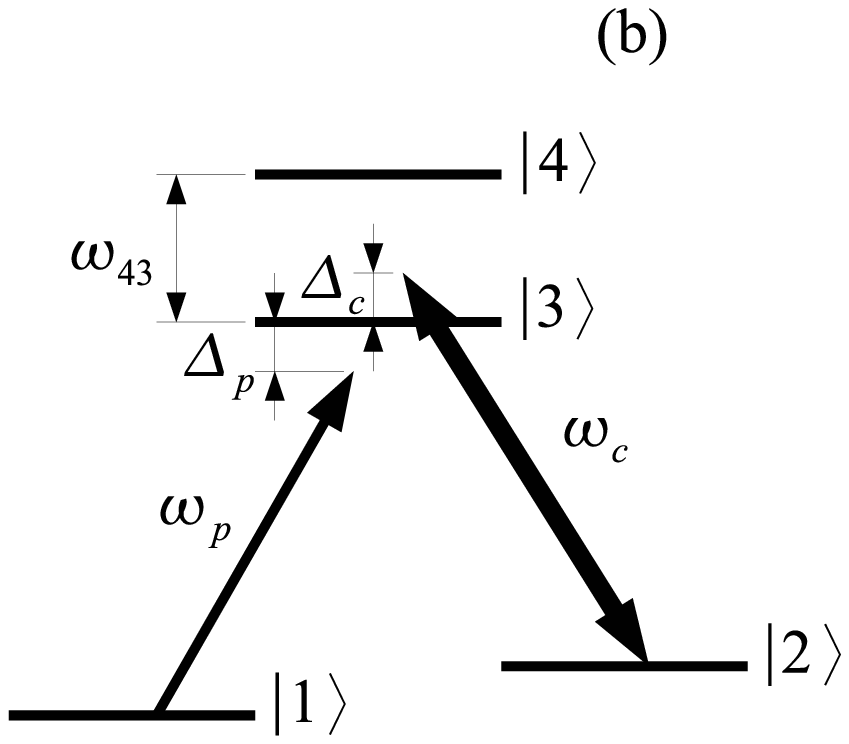}}
  \caption{(a) Energy level diagram of $^{87}$Rb. (b)
  Simplified four-level structure for EIT.}
  \label{fig:diagram}
\end{figure}

 We show our
theoretical analysis as follows. The energy level diagram of
$^{87}$Rb is shown in Fig.~\ref{fig:level}. The figure shows that
the excited levels of $^{87}$Rb are not singlets. The $5P_{3/2}$
level has 4 sublevels, and the $5P_{1/2}$ level has 2 sublevels. Two
of the sublevels $F=1$ and $F=2$ can form a $\Lambda$ structure for
EIT with the ground states $5S_{1/2}$. This structure can be
simplified to a four-level structure as shown in
Fig.~\ref{fig:setup}, in which there are two $\Lambda$ -type
structures: $|1\rangle-|3\rangle-|2\rangle$ and
$|1\rangle-|4\rangle-|2\rangle$ for EIT. If the energy difference
between $|3\rangle$ and $|4\rangle$ is not large enough, then these
two paths will interfere with each other, and the property of the
EIT will be changed, especially when the Doppler broadening is
considered. We make a detail analysis by using the master equation.
Considering a four level system with two fields $\omega_{p}$ and
$\omega_{c}$ as shown in Fig.~\ref{fig:setup}, we treat $\omega_{p}$
as the probe field, which is much weaker than the coupling field
$\omega_{c}$. The effective Hamiltonian of the system can be written
as
\begin{equation}
  \setlength\arraycolsep{5pt}
  H_{\rm int}=-\frac{\hbar}{2}\begin{pmatrix}
        0 & 0 & \Omega_{p3} & \Omega_{p4}\\
    0 & 2(\Delta_p-\Delta_c) & \Omega_{c3} &
    \Omega_{c4} \\
    \Omega_{p3} & \Omega_{c3} & 2\Delta_p & 0 \\
    \Omega_{p4} & \Omega_{c4} & 0 &
    2(\Delta_p-\omega_{43})
  \end{pmatrix},
  \label{eq:lig:H41}
\end{equation}
\vspace{-1em}

\noindent where $\Delta_p=\omega_p-\omega_{31}$,
$\Delta_c=\omega_{c}-\omega_{32}$, $\omega_{ij}$ is the frequency
difference between levels $\left|i\right>$ and $\left|j\right>$.
$\Omega_{pi}=\mu_{i1}E_p/\hbar$ and $\Omega_{ci}=\mu_{i2}E_c/\hbar$
are the Rabi frequencies of the fields with the corresponding
transitions,  $\mu_{ij}$ is the transition electronic dipole moment
of the $\left|i\right>\to\left|j\right>$ transition. Here we suppose
all $\Omega_{pi}$ and $\Omega_{ci}$ are real. When a cell filled
with buffer gas or coated with paraffin is used, the exchange of the
atoms can be ignored, therefore the decay between the ground states
is very small and can be ignored. The master equation for the atomic
density operator can be written as
\cite{Fleischhauer:2005,Chen:PRA:2008:013804}
\begin{figure}[tb]
  \centering
  \includegraphics[width=8cm]{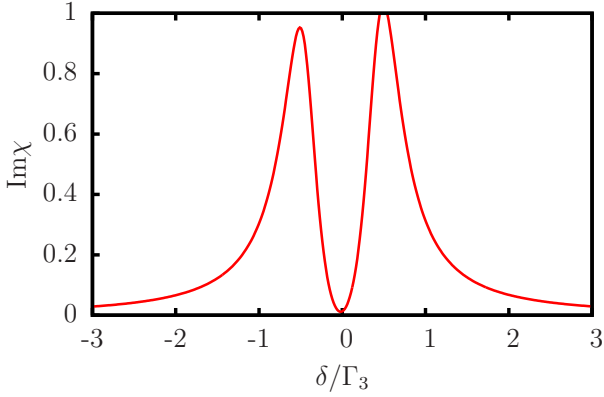}
  \caption{(Color online) Im[$\chi(\omega_{p})$] versus $\delta$ when no Doppler broadening
  is considered.}
  \label{fig:chi:nodop}
\end{figure}
\begin{eqnarray}
  \frac{d\rho}{dt}&=&\frac{1}{i\hbar}[H,\rho]+\frac{\Gamma_{31}}{
  2}(2\hat\sigma_{13}\rho\hat\sigma_{31}-\hat\sigma_{33}\rho-\rho\hat\sigma_{33})\nonumber\\
  &&+\frac{\Gamma_{32}}{2}(2\hat\sigma_{23}\rho\hat\sigma_{32}-\hat\sigma_{33}\rho-\rho\hat\sigma_{33})
  \nonumber \\
  &&+\frac{\Gamma_{41}}{
  2}(2\hat\sigma_{14}\rho\hat\sigma_{41}-\hat\sigma_{44}\rho-\rho\hat\sigma_{44})\nonumber\\
  &&+\frac{\Gamma_{42}}{
  2}(2\hat\sigma_{24}\rho\hat\sigma_{42}-\hat\sigma_{44}\rho-\rho\hat\sigma_{44})\nonumber\\
  &&+\frac{\gamma_{3\rm deph}}{2}(2\hat\sigma_{33}\rho\hat\sigma_{33}-\hat\sigma_{33}\rho-\rho\hat\sigma_{33})
  \nonumber \\
  &&+\frac{\gamma_{4\rm deph}}{
  2}(2\hat\sigma_{44}\rho\hat\sigma_{44}-\hat\sigma_{44}\rho-\rho\hat\sigma_{44})
  \nonumber \\
  &&+\frac{\gamma_{\rm2deph}}{
  2}(2\hat\sigma_{22}\rho\hat\sigma_{22}-\hat\sigma_{22}\rho-\rho\hat\sigma_{22}).
  \label{master}
\end{eqnarray}
We numerically solve Eq.~(\ref{master}) to obtain the linear
susceptibility $\chi(\omega_{p})$ concerned with $\omega_{p}$,
$\chi(\omega_{p})\propto
(\rho_{31}/\Omega_{p3}+\rho_{41}/\Omega_{p4})$, In the calculation,
we suppose $\mu_{31}=\mu_{41}=\mu_{32}=-\mu_{42}$
\cite{Chen:PRA:2008:013804}. The energy difference between
$5P_{3/2},F=1$ and $5P_{3/2},F=2$ is 157 MHz, which is about 26
times $\Gamma_{3}=\Gamma_{31}+\Gamma_{32}$ (about 6 MHz).
Substituting the data $\Omega_{p3}=\Omega_{p4}=0.001\Gamma_{3}$,
$\Omega_{c3}=-\Omega_{c4}=\Gamma_{3}$, $\Delta_{c}=0$ to
Eq.~(\ref{master}), we obtain Im[$\chi(\omega_{p})$] versus
$\delta=\Delta_{p}-\Delta_{c}$ as shown in Fig.~\ref{fig:chi:nodop}.
This figure shows that the existence of level $|4\rangle$ will
slightly affect the EIT spectrum: the EIT signal is not symmetric.

\begin{figure}[tb]
  \centering
  \includegraphics[width=8cm]{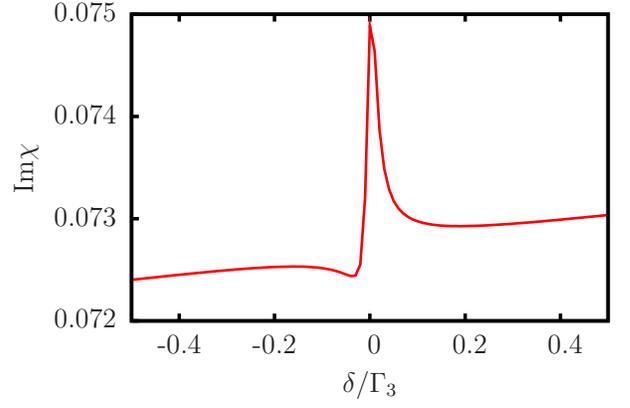}
  \caption{(Color online) Imaginary part of the susceptibility with Doppler integration.}
  \label{fig:chi}
\end{figure}

Following, we consider the effect of Doppler broadening. The
distribution function of the frequency shift with respect to the
center frequency $f_0$ can be simplified as
\begin{equation}
  P(f)\propto \exp \left( -\frac{m\lambda_{0}^{2}(f-f_{0})^{2}}{2kT} \right),
  \label{eq:dist}
\end{equation}
where $k$ is the Boltzmann constant,  $T$ is the temperature. $f$ is
the frequency, $m$ is the mass of $^{87}$Rb, and $\lambda_{0}$ is the wavelength
of the corresponding transition. Substituted the data of $^{87}$Rb
and $T=320$ K into Eq.~(\ref{eq:dist}), the imaginary part of the
susceptibility after Doppler integration is shown in
Fig.~\ref{fig:chi}.
This figure clearly shows that the EIT signal has been ruined
completely by the Doppler broadening. Instead of the transparency at
$\delta=0$, there is an enhanced absorptive peak. This absorptive
peak is very small compared with the background, therefore we have
not observed it in the experiment yet. The disappearance of the
transparency window makes the atomic ensemble opaque to the photon.
Therefore coherent photons can not be generated.

\begin{figure}[tb]
  \centering
  \includegraphics[width=8cm]{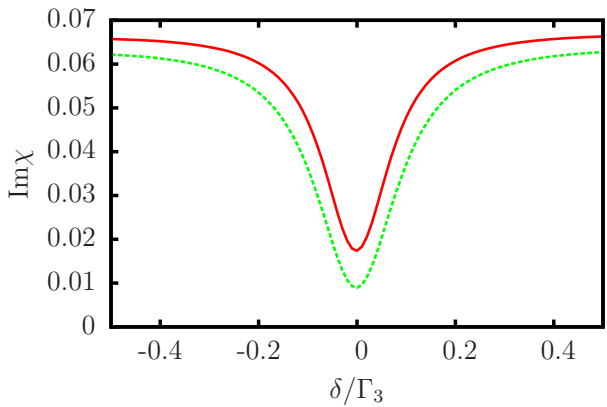}
  \caption{(Color online) Im[$\chi(\omega_{p})$] versus $\delta$ when decay caused by the atom
  exchange is considered. Red solid line is the result of Eq.~(\ref{eq:ma1}),
  and green dashed line is the result of Eq.~(\ref{eq:ma2}).}
  \label{fig:decay}
\end{figure}

When a normal cell is used, the exchange of the atoms should be
considered, the decay time between the ground state is short.
The atoms leaving and entering the
light beam can be considered as an effective decay between the ground
states, the master equation for a normal cell can be denoted as
\begin{equation}
  \frac{d\rho}{dt}=M-r\rho+\frac{r}{2}(\hat\sigma_{11}+\hat\sigma_{22}),
  \label{eq:ma1}
\end{equation}
or
\begin{eqnarray}
  \frac{d\rho}{dt}&=&M+\frac{\gamma}{2}(2\hat{\sigma}_{12}\rho\hat{\sigma}_{21}-\hat{\sigma}_{22}\rho-\rho\hat{\sigma}_{22})\nonumber\\&&+\frac{\gamma}{2}(2\hat{\sigma}_{21}\rho\hat{\sigma}_{21}-\hat\sigma_{11}\rho-\rho\hat{\sigma}_{11}),
  \label{eq:ma2}
\end{eqnarray}
where $M$ is the right side of Eq.~(\ref{master}), $r$ is the
exchange rate of atoms. $\gamma$ is the
effective decay between the ground states caused by the exchange of
atoms. Equation~(\ref{eq:ma1}) gives a direct description of the
atoms leaving and entering the field, and Eq.~(\ref{eq:ma2}) shows
the effective decay between the ground states caused by the exchange
of atoms. Although these two descriptions are different, they show
the similar effect on the EIT caused by the exchange of atoms. The
imaginary parts of the susceptibility with Doppler broadening at
$r=0.01\Gamma_{3}$ and $\gamma=0.01\Gamma_{3}$ are shown in
Fig.~\ref{fig:decay}. The figure shows that both of the simulations
have the similar results: the decay caused by the exchange of atoms
can enhance the EIT of the system. This result agrees with our
experimental result of the EIT on the D2 transition of $^{87}$Rb,
and also agrees with the experimental result about generation of the
photon pairs we observed in the experiment.

\begin{figure}[tb]
  \centering
  \subfigure{\label{fig:rsnt}
  \includegraphics[width=4cm]{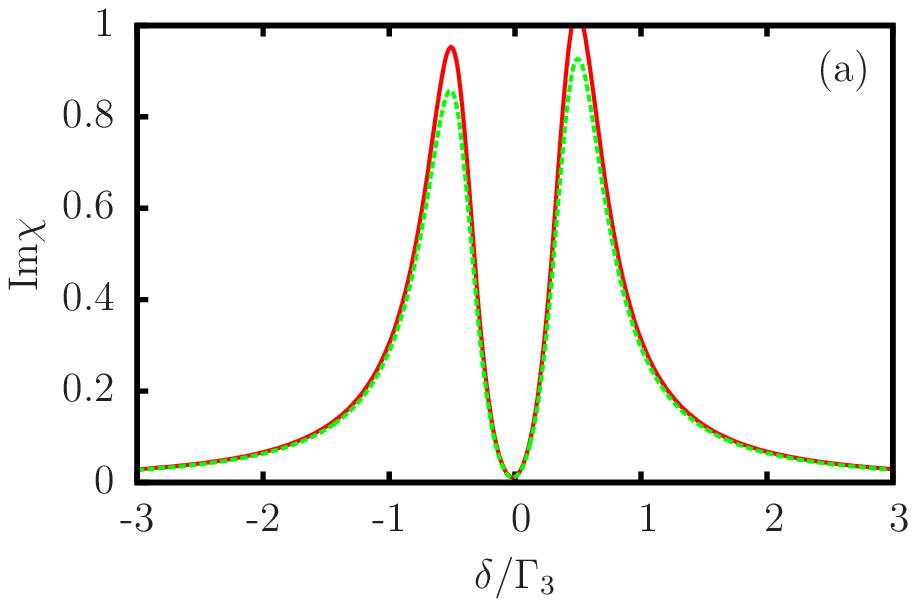}}\subfigure{\label{fig:dtn}\includegraphics[width=4cm]{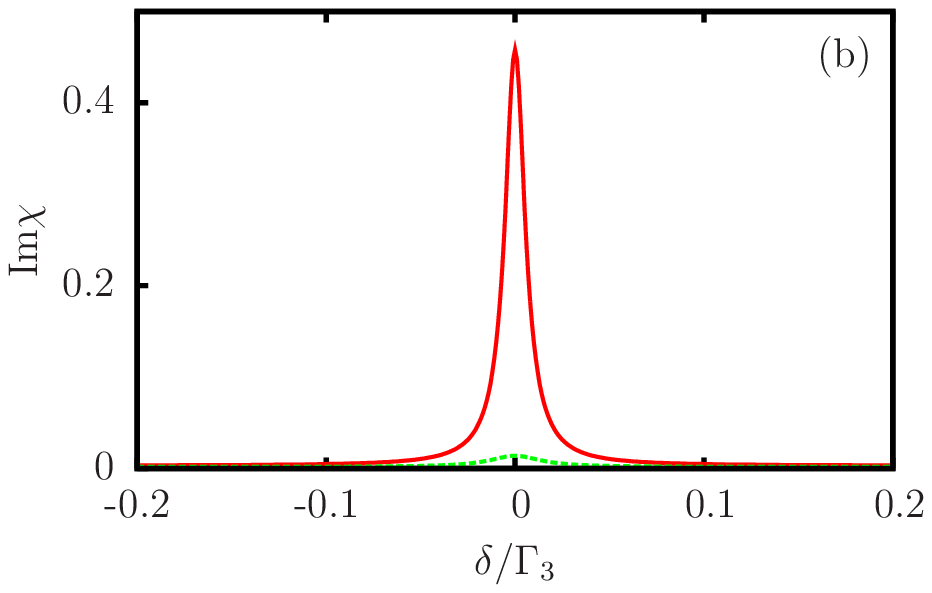}}
  \caption{(Color online) Comparison of EIT with and without decay between the ground states.
  (a) The coupling is resonant with the $|2\rangle\to|3\rangle$ transition. (b)
  The coupling is at the center of the $|2\rangle\to|3\rangle$ and
  $|2\rangle\to|4\rangle$ transitions. Red solid line: without decay; Green
  dashed line: with decay.
  }
  \label{fig:cmp}
\end{figure}
The reason why the decay can enhance the EIT is that the decay makes
the EIT signal reduced very quickly as the increment of detuning of
the coupling. Therefore the interfere between the two EIT paths is
small enough and the EIT can be preserved even the Doppler
broadening exists. To support this point, we show the numerical
result of the comparison of the EIT with and without the decay,
which correspond to the cases in which the coupling is resonant with
the $|2\rangle\to|3\rangle$ transition and is at the center of the
$|2\rangle\to|3\rangle$ and $|2\rangle\to|4\rangle$ transitions. The
results are shown in Fig.~\ref{fig:cmp}. Figure \ref{fig:rsnt} shows
that the decay does not affect the EIT too much when the coupling is
resonant with a transition. Figure \ref{fig:dtn} shows that when the
coupling is detuned from the transition, the interference of the two
paths can cause a large absorptive peak at $\delta\approx0$, which
makes the disappearance of EIT after considering Doppler broadening;
The existence of the decay between the ground states makes the
absorptive peak weakened very quickly as the detuning of the
coupling is increased, this makes the EIT preserved even Doppler
broadening exists.

In the case of the D1 transition of the $^{87}$Rb, because the
energy split of $5P_{1/2}$ is large enough, the EIT signal will
always exists after integration of Doppler broadening. That is the
reason why the work reported in
Ref.\cite{Wal:2003:196,Eisaman:PRL:2004:233602,Eisaman:N:2005:837,
manz:040101}can generate non-classical photon pairs
successfully.

In conclusion. We make a detail theoretical analysis on the EIT at the D2
transition of the hot $^{87}$Rb, which shows the long decay time between the
ground states will ruin the EIT. This analysis shows a probable 
reason why a rubidium cell with long decay time is not a useful candidate
for preparation of a non-classical photon pair via the D2
transition. The simulations agree well with the experimental
results.  We believe our
find can give a very useful instruction to such kind of research.

  We thank Wei Jiang for some useful discussions. This work is funded by National
  Fundamental Research Program (Grant No.  2006CB921900, 2009CB929601), National
  Natural Science Foundation of China (Grants No. 10674126, No. 10874171), the
  Innovation funds from Chinese Academy of Sciences, Program for NCET, and
  International Cooperate Program from CAS.


\end{document}